\documentclass[aps,pra,twocolumn,superscriptaddress]{revtex4-1}

\usepackage{amsmath}
\usepackage{graphicx}
\usepackage[export]{adjustbox}
\usepackage{hyperref}
\usepackage{color}
\usepackage{mathrsfs}
\usepackage{lipsum}
\usepackage{setspace}
\usepackage{amsfonts}
\usepackage{amssymb}
\usepackage{calrsfs}

\usepackage{amssymb}
\usepackage{amsfonts}
\usepackage{amssymb}
\usepackage{epstopdf}
\usepackage{comment}
\usepackage{ tipa }
\usepackage{braket}
\usepackage{physics}
\usepackage{float}

\newcommand{\be}{\begin{eqnarray}}
\newcommand{\ee}{\end{eqnarray}}

\begin{document}

\title{Visualizing Strange Metallic Correlations in the 2D Fermi-Hubbard Model with AI}

\author{Ehsan Khatami}
\email[]{ehsan.khatami@sjsu.edu}
\affiliation{Department of Physics and Astronomy, San Jos\'{e} State University, San Jos\'{e}, CA 95192, USA}
\author{Elmer Guardado-Sanchez}
\affiliation{Department of Physics, Princeton University, Princeton, NJ 08544,USA}
\author{Benjamin M. Spar}
\affiliation{Department of Physics, Princeton University, Princeton, NJ 08544,USA}
\author{Juan Felipe Carrasquilla}
\affiliation{Vector Institute, MaRS  Centre,  Toronto,  Ontario,  M5G  1M1,  Canada}
\author{Waseem S. Bakr}
\affiliation{Department of Physics, Princeton University, Princeton, NJ 08544,USA}
\author{Richard T. Scalettar}
\affiliation{Department of Physics, University of California, Davis, CA 95616, USA}

\date{\today}

\begin{abstract}
Strongly correlated phases of matter are often described in terms of
straightforward electronic patterns. 
This has so far been the basis for 
studying the Fermi-Hubbard model realized with ultracold atoms. Here, 
we show that artificial intelligence (AI) can provide an unbiased 
alternative to this paradigm for phases with subtle, or even unknown, 
patterns. Long- and short-range spin correlations spontaneously emerge 
in filters of a convolutional neural network trained on snapshots of 
single atomic species. In the less well-understood strange metallic 
phase of the model, we find that a more complex network trained on 
snapshots of local moments produces an effective order parameter for 
the non-Fermi-liquid behavior. Our technique can be employed to 
characterize correlations unique to other phases with no obvious order 
parameters or signatures in projective measurements, and has 
implications for science discovery through AI beyond strongly 
correlated systems.
\end{abstract}

\maketitle

%%%%%%%%%%%%%%%%%%%%%%%%%%%%%%%%%%%%%%%%%%%%%%%%%%%%%%%%%%%%%%%%%%%%
\section{Introduction}
%%%%%%%%%%%%%%%%%%%%%%%%%%%%%%%%%%%%%%%%%%%%%%%%%%%%%%%%%%%%%%%%%%%%

Strongly correlated phases of matter are often described in terms of
relatively simple real space order parameters, 
which are theoretically understood using 
Landau symmetry-breaking theory~\cite{wen2004quantum}. For instance, 
ferromagnetism on a square lattice involves a uniform arrangement where the 
electrons' spins align and create a magnetic state with a wavevector 
${\bf q=0}$. Antiferromagnetism, slightly more complex, is revealed by a 
 ${\bf q} =$ {\boldmath $\pi$} alternation of the electrons' spin 
 state on two sublattices. These choices, and incommensurate
(spiral) order which bridges them at general ${\bf q}$, can be
characterized in a unified way through the magnetic structure factor, 
$S({\bf q})$, and further generalized to include time-domain patterns 
via the dynamic susceptibility, $\chi({\bf q},\omega)$.
Similar statements apply to charge density wave and other phases
involving diagonal long-range order.  

While many of our theoretical and experimental probes of interacting 
quantum systems have been constructed with coupling to these patterns 
in mind, there is an increasing realization that the most interesting
strongly correlated phases might not be immediately accessible via such
observables. Cuprate and iron pnictide superconductors, which combine closely 
entwined conventional phases with well-established order parameters, and much
less well-understood non-Fermi liquid (NFL) or pseudogap phases with so far
``hidden orders" are 
examples~\cite{j_orenstein_00,c_varma_02,s_sachdev_16}, as is the zoo of orbital 
ferromagnetism, superconductivity, and Mott insulating behavior in twisted 
bilayer graphene~\cite{y_cao_18,y_cao_18b}.
The community of strongly correlated quantum systems is thus faced with
the challenge of developing new means of identifying complex phases.

Here, we introduce an unbiased approach in which artificial intelligence (AI) is used to extract hidden 
features from raw images of quantum many-body systems.
We test our approach using projective measurements 
on a two-dimensional (2D) Fermi-Hubbard model, obtained through quantum 
gas microscopy of ultracold fermionic atoms in an optical lattice.
We find that filters of a convolutional neural network (CNN), trained to recognize 
snapshots of fermions, capture features at different densities that have clear 
interpretation in terms of short- and long-range magnetic correlations. 
We further show that a more complex CNN can produce an 
effective order parameter for the NFL phase, based on the interplay of multiple 
types of density fluctuations, reflecting the more enigmatic 
nature of the correlations in this phase.

In the experiment, the 2D Fermi-Hubbard model is realized using a spin-balanced 
mixture of the first and third lowest energy states of $^6$Li loaded into a square 
optical lattice. We work at a magnetic field of 615 G in the vicinity of the 
Feshbach resonance near 690 G, which gives a scattering length of 
1056(10)$\mathrm{a_0}$, where $\mathrm{a_0}$ is the Bohr radius 
The lattice depth is $7.25(2) E_R$, where $E_R$ is the lattice 
recoil energy and $E_R / h = 14.66$ kHz. For these parameters we obtain 
$t/h=850(20)$~Hz and $U/t = 8.0(1)$. Here, $t$ and $U$ are the nearest-neighbor 
hopping matrix element and the strength of onsite repulsive interaction, respectively, 
in the Hubbard model (see Appendix A).

Using quantum gas microscopy techniques~\cite{p_brown_17}, we image the atoms 
in the lattice with single site resolution with a fidelity of 98\%. When a 
fluorescence image is taken, atoms on doubly occupied sites undergo light-assisted 
collisions and appear empty. An image taken this way allows us to extract the 
local moment on each site. Alternatively, we can apply a short pulse of resonant 
light prior to taking an image to eject atoms of one of the two hyperfine states. 
This allows us to measure the single component density of the remaining hyperfine state.

Our lattice beams produce a harmonic trapping potential, which if uncompensated 
leads to significant variations of the local density. To study regions of uniform 
density, we flatten the potential using light shaped using a spatial light 
modulator~\cite{p_brown_18}. In the subsequent analysis, we work with a flattened 
region of $20\times 20$ lattice sites.

\begin{figure}
\centering
\includegraphics*[width=3.3in]{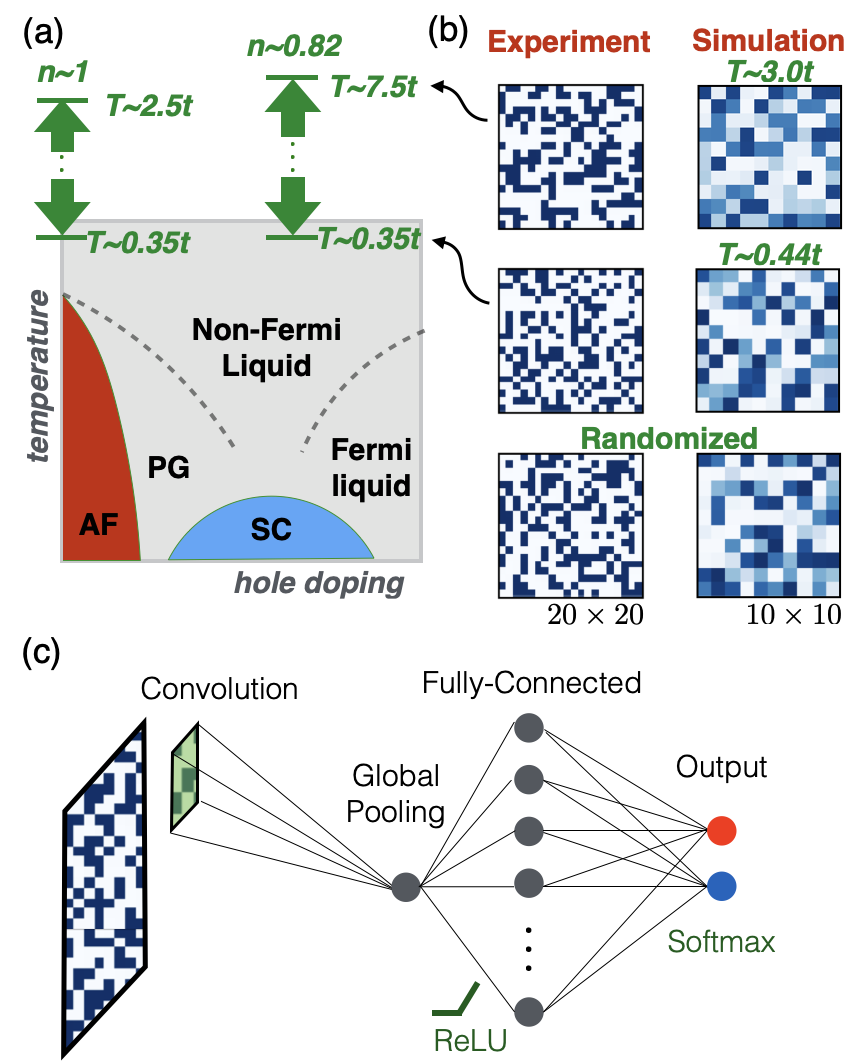}
\caption{{\bf Phase diagram, sample snapshots, and CNN architecture.}
(a) Schematic phase diagram of cuprate high-temperature superconductors
in the space of temperature and hole doping. AF, PG and SC stand for 
antiferromagnetic, pseudogap, and superconducting phases, respectively. 
(b) Sample experimental (left) and determinant quantum Monte Carlo (right) snapshots 
taken for $U=8t$ at the density $n \sim 0.82$ at different temperatures. The 
bottom row shows the lower-temperature snapshots whose pixels have 
been randomly shuffled, i.e., ``fake" snapshots. 
(c) The main convolutional neural network architecture used in this study.
It contains a convolutional layer with one filter and one feature map
followed by a global pooling layer, a hidden layer with eight 
fully connected neurons, and an output softmax layer with two neurons, each 
associated with a temperature limit. We use the rectified linear unit (ReLU) as 
the activation function in all but the output layer. 
In our experiments we observe that the presence of the fully connected layer 
accelerates the training of the neural network.\label{fig:sample}}
\end{figure}

Figure~\ref{fig:sample} shows two randomly chosen samples of binarized 
occupancy snapshots at an average density of $n=0.82(2)$
at two extreme temperatures of $T\sim U \sim 8\,t$ and $T\sim0.35\,t$. 
These parameters place us within the NFL region of a typical cuprate 
phase diagram [see Fig.~\ref{fig:sample}(a)]. Thermometry is performed using 
averages of various correlation functions taken over such 
snapshots~\cite{p_brown_18}.

\begin{table}[t]
\begin{tabular}{cccccccc}
\hline
\hline
$n=\to$ & $0.97$ & $0.835$ & $0.82$ & $0.735$ & $0.70$ & $0.64$ & $0.58$  \\
 \hline
 Spin-up/spin-down & $402$  &  & $216$ &  &  &  & \\
 Singles & $201$ & $281$ & $5023$ & $290$ & $342$ & $281$ & $330$ \\
 \hline
 \hline
\end{tabular}
\caption{Number of available experimental snapshots at $T\sim 0.35$ for different densities.}
\end{table}

The increasingly large number of snapshots
taken in quantum gas microscope experiments in various regions of the parameter space
lends itself to data-driven approaches for science discovery, such 
as the enlisting of AI (see Table I for the number of snapshots used in this study). In fact, early implementations of 
machine learning techniques for the study of quantum many-body systems
demonstrated great potential~\cite{Carleo2016,j_carrasquilla_16,k_chng_17,d_deng_16,e_vanNieuwenburg_17,y_zhang_17a,p_zhang_18}.
Recent applications to experimental data have 
directly led to the discovery of new physics~\cite{f_zhang_18,a_bohrdt_18,b_rem_19,g_torlai_19,a_samarakoon_20}, modeling of their distribution~\cite{c_casert_20}, 
or the optimization of experimental processes~\cite{p_wigley_16,r_lewis_19}, 
including those related to quantum gas microscopy.

CNNs offer an ideal platform for the detection of patterns in the experimental snapshots. 
Not only can they efficiently compress the information in images and use them for 
classification, but also their trained filters provide a window into the relevant 
features observed~\cite{m_zeiler_14}. Figure~\ref{fig:sample}(c) shows the main CNN architecture 
we have used. After labeling them according to their temperature, hundreds of 
snapshots taken at the extreme temperatures along with their labels are provided to the CNN for training. 
During the training, the network adjusts its free parameters to minimize the difference
between given labels and its prediction (see Appendix B). 
The convolutional layer in our CNN interacts directly with the input snapshots and,  
therefore, examining the filter after the completion of training can teach us 
about the most important feature the network has picked up. 

%%%%%%%%%%%%%%%%%%%%%%%%%%%%%%%%%%%%%%%%%%%%%%%%%%%%%%%%%%%%%%%%%%%%
\section{Results}
%%%%%%%%%%%%%%%%%%%%%%%%%%%%%%%%%%%%%%%%%%%%%%%%%%%%%%%%%%%%%%%%%%%%

\begin{figure}
\centerline {\includegraphics*[width=3.3in]{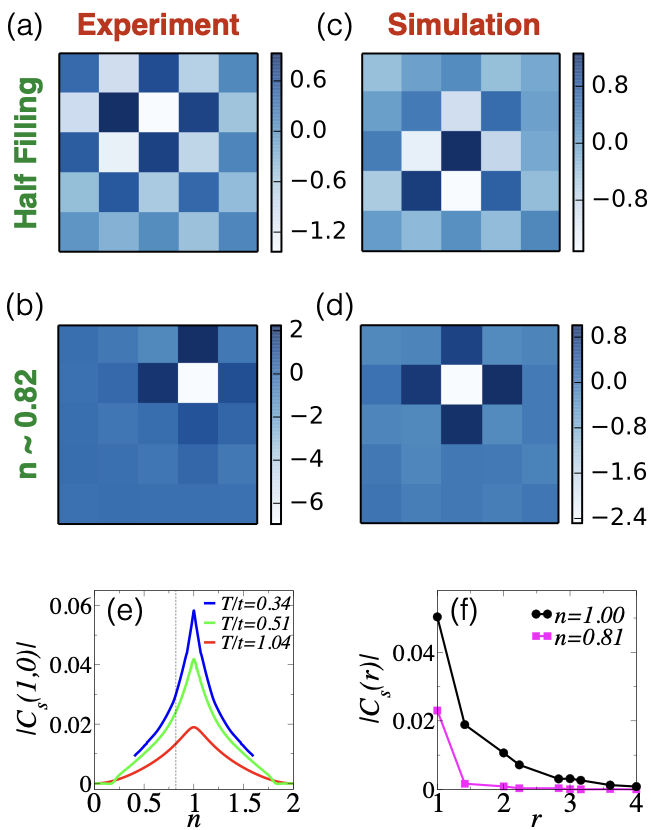}}
\caption{{\bf Analysis of single-species snapshots using CNNs with one filter.} 
Sample $5\times 5$ trained filters for (a) $n \sim 1$ and (b) $n = 0.82$.
The CNN architecture is shown in Fig.~\ref{fig:sample}(c). 
The testing accuracies are between 91\% and 96\%. The visual pattern 
in (a) is consistent with recognizing long-range antiferromagnetic (AF) order near half filling. The
filter in (b) indicates a pattern capturing short-range AF correlations. 
(c, d) Similar sample filters evolved from training 
runs using determinant quantum Monte Carlo (DQMC) simulations at $T=0.1t$ and $0.44t$, with testing 
accuracies of 91\% and 68\%, respectively. 
(e, f) Theory data for the nearest-neighbor
spin-spin correlation for $U=8t$ vs~density at different temperatures (numerical linked
cluster expansion), and
vs~distance for $n = 0.81$ and $n = 1.00$ at $T=0.44t$ (DQMC). These results illustrate that
AI can capture the correct trends in magnetic behavior of the Hubbard model, 
and that the trained filters carry a clear physical interpretation. 
\label{fig:5x5}}
\end{figure}

Figure~\ref{fig:5x5}(a) shows a sample $5\times 5$ filter for a CNN that 
is trained to distinguish experimental 
snapshots of a single species of fermions at the highest temperature ($T\sim2.5t$) from those
at the lowest temperature ($T\sim0.35t$) when $n\sim 1$. If we expect mostly random behavior at high 
temperature, of the same order as the largest energy scale in the system, 
the features that spontaneously develop in the filters during training will most 
likely represent patterns found in the low-temperature snapshots. We find that the CNN 
consistently makes the distinction with more than 91\% accuracy, and 
it does so using filters showing a distinctive pattern indicative of long-range 
antiferromagnetic (AF) correlations.

Training the CNN using similar snapshots obtained for $n=0.82$ at $T\sim7.5t$ and $0.35t$
results in filters that reflect a shorter-range anticorrelation between 
neighboring fermions of the same species [see Fig.~\ref{fig:5x5}(b)]. 
The nearest-neighbor checkerboard pattern
emerging in the filters is consistent with the fact that the correlation length 
in the NFL region is about one lattice spacing~\cite{j_tranquada_07}. 
We find that this feature appears at different locations in the filter for 
different training runs, which points to a redundancy: on average the filter must 
reflect the translational symmetry of the underlying system.
These findings suggest that the network effectively uses the strength 
of AF correlations as a measure for classifying snapshots of a single species
of fermions. Figures~\ref{fig:5x5}(e) and ~\ref{fig:5x5}(f) 
show that the density, distance, and
temperature dependence of the magnetic correlations of the model, 
$C_s({\bf r})$ (see Appendix A), which are calculated here on a $10\times 10$ cluster
using the determinant quantum Monte Carlo (DQMC) method~\cite{r_blankenbecler_81}, 
or in the thermodynamic limit using the numerical linked cluster expansion 
(NLCE)~\cite{M_rigol_06,E_khatami_11b}, support this observation.

Quantum Monte Carlo simulations also provide a platform to corroborate  
these findings. However, except in one spatial dimension, these 
simulations cannot provide projective measurements in the density 
basis. Instead, theory ``snapshots" can be constructed via expectation 
values of local charge or spin density using instances of auxiliary 
field variables during a simulation; for example, the $i$th 
pixel of a spin-up DQMC snapshot is  
$\left<\hat{n}_{i\uparrow}\right>_{h}=1-\mathcal{G}_{ii\uparrow}(h)$, 
where $\mathcal{G}_{ii\uparrow}(h)$ is the $i$th diagonal element of the 
spin-up equal time Green's function matrix for the auxiliary field instance $h$. 
We perform the simulations for a $10\times 10$ site Hubbard system with 
$U=8t$ at several average densities and temperatures (see Appendix D). 

At high temperatures, 
of the order of $3t$, we find that density snapshots are fuzzy with no clear 
empty sites; mostly fluctuations about an average background density can be seen.
This fuzziness is less of a concern for single-species snapshots 
[see Fig.~\ref{fig:sample}(b)], although they too lose their pixelated character at 
higher temperatures. For this reason, to eliminate 
fuzziness as an obvious feature for the CNN to learn, instead of high-temperature
snapshots, we use low-temperature
images whose pixels have been randomly shuffled, effectively destroying any physical 
correlations. In the following, we refer to the 
latter as {\it fake} (as opposed to {\it real}) snapshots.

Figures~\ref{fig:5x5}(c) and \ref{fig:5x5}(d) show sample filters from
training experiments using theory snapshots of single species at half filling and $n=0.82$. 
Despite reduced accuracies of about 68\% for the latter, which we
believe is due to the exacerbation of the issue with the nonprojective nature of simulated images at this
density, we find that the trained features 
are in excellent agreement with those obtained with quantum gas microscope 
snapshots. Together, they demonstrate that relevant spin correlations can be
captured in an unbiased fashion through CNNs.

Studies of the origin of the NFL behavior, 
a central question in any theory of high-temperature superconductivity~\cite{s_sachdev_10},
have for decades been focused on its possible connections to the order parameter fluctuations of 
a magnetic quantum critical 
point~\cite{j_hertz_76,s_sachdev_92a,a_millis_93,g_stewart_01,h_lohneysen_07,s_sachdev_10,a_fitzpatrick_13}.
Here, we are in a position to ask whether any such fluctuations manifest 
themselves in charge correlations too, and to what extent they can be inferred 
from the other type of snapshots available in the experiment, those of local moments.

A similar analysis using images at the two extreme temperatures, however,
is largely affected by the abundance of doubly occupied sites at $T\sim 7.5t$, 
and their lack of representation in the snapshots of local moments. 
Upon lowering the temperature to $T\sim 0.35t$, the fraction of doubly occupied 
sites at 18\% doping reduces roughly by a factor of 4, from 12\% to about 3\%~\cite{E_khatami_11b},
providing the CNN again with an obvious feature with which to perform classification.
Removing this bias by randomly populating pixels to create ``fake"
replacements for high-temperature snapshots in the training, brings 
the accuracy down dramatically when $n=0.82$.

\begin{figure}
\centerline{\includegraphics*[width=3.3in]{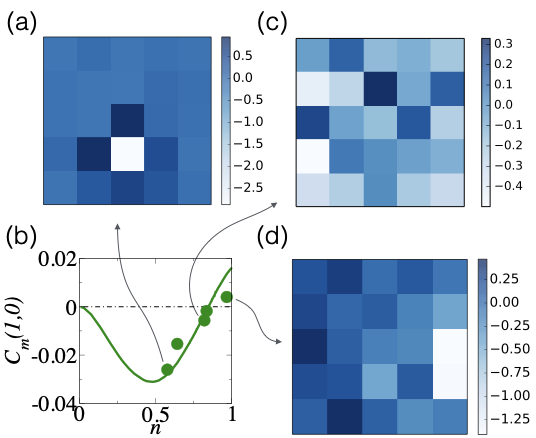}}
\caption{{\bf Analysis of local moment snapshots using CNNs with one filter.} 
Representative $5\times 5$ filters for runs at (a) $n = 0.58$, 
(c) $n =0.82$, and (d) $n = 0.97$ using the same 
CNN architecture as one used for snapshots of single species. (b) 
The nearest-neighbor local moment correlation function from DQMC (solid 
line) on an $8\times 8$ cluster at $T=0.3t$ and from the experiment (circles) at 
similar low temperatures. The testing accuracy reaches (a) $89\%$, (c) $62\%$,
and (d) $60\%$.\label{fig:ExptLM}}
\end{figure}

Figure~\ref{fig:ExptLM} shows that the accuracy of CNNs trained on snapshots of 
local moments largely depends on the strength of short-range correlations between the moments.
The largest accuracies (almost $90\%$) are typically achieved near quarter 
filling, where the correlations are the most negative. Patterns observed in 
filters trained in this region are also consistent with the anticorrelation of neighboring moments
[Fig.~\ref{fig:ExptLM}(a)]. The accuracy drops to around $60\%$ at $n= 0.82$, 
near the zero crossing of the correlator, which is shown in Fig.~\ref{fig:ExptLM}(b). 
Typical trained filters do not display any immediately recognizable patterns 
either [see Fig.~\ref{fig:ExptLM}(c)]. Near half filling, the correlations
between local moments are positive due to the bunching of holes and 
doubly occupied sites~\cite{l_cheuk_16}. Here, we find that despite the relatively 
low accuracies ($\lesssim 65\%$), trained filters often do reflect the
bunching of empty sites [Fig.~\ref{fig:ExptLM}(d)].

\begin{figure*}
\centering{\includegraphics*[width=\textwidth]{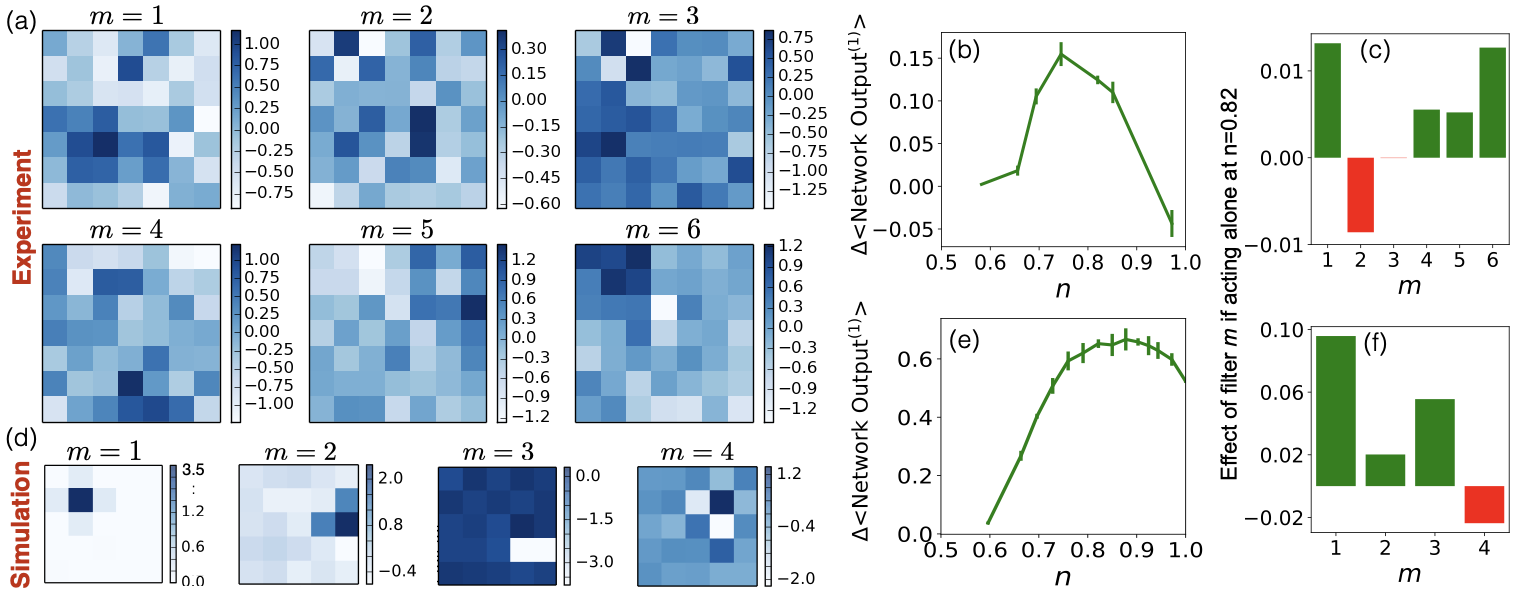}}
\caption{{\bf Analysis of local moment snapshots using CNNs with multiple filters.} 
(a) The six filters of a trained CNN. Training is performed with the 5023 
experimental local moment snapshots taken at $n= 0.82$ and $T\sim 0.35$. 
The testing accuracy remains at $62\%$ with less than $1\%$ variation over the last 20 epochs (see Appendix B).
(b) The difference in the average network output for $T\sim 0.35t$ real and fake snapshots 
as a function of the density when all six filters are present: 
$\Delta\left<\text{Network Output}^{(1)} \right>\equiv \left<\text{Network Output}^{(1)}  
({\bf X}^{\text{real}}) \right>-\left< \text{Network  Output}^{(1)}  ({\bf X}^{\text{fake}}) \right>$. 
Superscript $(1)$ indicates the value at the output neuron responsible 
for real low-temperature snapshots (see Appendix B).
This quantity indicates roughly the percentage of the output attributable 
to factors other than the density.
(c) Similar to (b) at $n= 0.82$ when the CNN has access to one filter at a time.  
(d) Four representative filters of a CNN with sixteen $5\times 5$ 
filters trained using DQMC snapshots of local moments (see Appendix D).
(e, f) Same as (b) and (c), but obtained using the CNN in (d).\label{fig:Expt6filter}}
\end{figure*}

The comparison of Fig.~\ref{fig:ExptLM}(c) ($n=0.82$) with Figs.~\ref{fig:ExptLM}(a)
and ~\ref{fig:ExptLM}(d) ($n=0.58, 0.97$) makes it clear  
that snapshots of local moments in the NFL region of the Hubbard 
model do not contain a single dominant ordering pattern and that a more 
advanced treatment may be necessary to capture the physics.
In Fig.~\ref{fig:Expt6filter}, we show results of a training with a CNN, 
modified to include six $7 \times 7$ filters in its convolutional layer (see Appendix B).
The bigger data set we have available for snapshots of local moments at this density 
allows us to experiment with different filter sizes and numbers of filters.
We find that including more than one filter in the CNN improves the best accuracies 
only marginally in this case, up to around 65\%, and having too 
many and/or much larger filters can still result in overfitting. 
We also find that using a deeper CNN with two convolutional layers does 
not significantly improve the accuracy.

Figure~\ref{fig:Expt6filter}(a) shows six filters of a sample CNN trained on the 
local moment snapshots. Their fuzzy patterns offer some insight
into possible spatial arrangements of local moments at low temperatures. As we see below, 
patterns in filters $m=1, 4, 5$, and $6$ are more frequently associated 
by the network with real low-temperature snapshots at this filling, whereas patterns 
in filter $m=2$ are more frequently associated with fake snapshots.

By transferring the knowledge of the CNN to other densities, we find that 
the network is the most sensitive to correlations around the NFL region. 
Figure~\ref{fig:Expt6filter}(b) shows the difference in probabilities 
that a snapshot and its fake counterpart are categorized as belonging to the NFL region,
effectively eliminating density itself as a factor in the signal. We find this 
quantity to be maximal in the vicinity of $n= 0.8$, suggesting that 
the CNN as a whole is in fact focusing on local moment correlations more unique 
to the NFL region and slightly lower densities.

While the contribution of individual filters to the CNN's decision making cannot be completely 
isolated, we can study what the network output would be if each filter were to act alone (see Appendix B). 
Figure~\ref{fig:Expt6filter}(c) shows this quantity averaged over samples at $n=0.82$, 
after subtracting the value for the corresponding fake snapshot,
for each of the six filters shown in Fig.~\ref{fig:Expt6filter}(a). The results suggest that
filters 1 and 6, if acting alone, would have the largest effect on the decision 
making at this average density, followed by filters 2, 4, and 5, while 
filter 3 plays almost no role at all. The negative value for filter 2 indicates that 
the network signal is larger on average for fake snapshots in that case.

Using DQMC, we verify that similar trends can be observed in simulated 
snapshots of local moments. However, unlike with the experimental snapshots, 
here, we find that the accuracy generally increases with increasing 
the number of filters in the CNN, while increasing the filter size does not 
necessarily improve the performance. We attribute these to the fundamental difference 
between the two types of snapshots (projective vs nonprojective). 
Figure~\ref{fig:Expt6filter}(d) highlights a representative sample of $5\times 5$ 
filters of a CNN with 16 such filters, trained on simulated snapshots 
reaching to an accuracy of 87\% (see Appendix C). They appear to measure a variety of 
short-range correlations to assist the network in making decisions.
Figure~\ref{fig:Expt6filter}(e) shows the overall signal of the CNN for 
correlations unique to the NFL phase, plotted across densities. It has 
a broad peak around the NFL region. As shown in Fig.~\ref{fig:Expt6filter}(f), 
patterns in the first three filters seem to be mostly associated with real snapshots in 
the NFL region, while the pattern in the $m=4$ filter is mostly associated with the fake snapshots.
We find that including the information about doubly occupied sites, i.e., using full 
density snapshots, generally improves the diversity of features seen in the trained 
filters while yielding the same basic trends.

\section{Discussion}

Using specially designed artificial neural networks,
we have developed algorithms for extracting organizing patterns of
correlated particles from raw quantum many-body data 
in an unbiased fashion and without any prior theoretical 
knowledge. When applied to the snapshots of one of the two species of
fermionic atoms in a 2D optical lattice, our approach yields patterns indicative
of long- and short-range AF correlation near and away from the commensurate filling, consistent with theory.
We show that these features can be reproduced using nonprojective measurements from DQMC simulations.

Our analysis provides a window into the signatures of the NFL phase, one 
of the most mysterious and theoretically challenging phases in correlated 
electron systems, in snapshots of local moments from the quantum gas microscope. We show that in this case, a more
complex neural network can be constructed and trained to be sensitive to 
correlations specific to the strange metallic phase. A similar 
analysis of snapshots with information about both species of particles in future 
experiments~\cite{j_koepsell_20} may further reveal the interplay between spin and 
charge fluctuations in this region.

Nonlinearities in the neural network model 
make the interpretation of features seen in the filters vis-\`{a}-vis correlations 
in the physical snapshots challenging since the knowledge of the network can be 
divided in nontrivial ways among its different components. An early example 
of this was the surprisingly successful classification of 
snapshots of the Ising lattice gauge theory at $T=0$ and $T=\infty$, despite the 
lack of an order parameter, using CNNs with multiple filters~\cite{j_carrasquilla_16}. 
The procedure we have introduced here overcomes aspects of this challenge. 
We point out that the nonlinearities are key for the success of our CNNs; 
for example, the principal component analysis~\cite{pca}, a linear unsupervised learning method,
largely fails to draw any meaningful distinction between sets of snapshots (see Appendix E).

The techniques developed in this work for the AI-assisted feature extraction 
in projective measurements can be adapted to peek into other 
mysterious phenomena for the Fermi-Hubbard model, such as
the pseudogap phase~\cite{c_chiu_19}, or the magnetic polaron which has been 
observed closer to half filling~\cite{j_koepsell_19}. They can also 
be employed to study other microscopic models of correlated
systems. Our work paves the way for AI-related studies that go beyond 
mere categorization and the quest for gaining more predictive power 
and focus instead on the inner workings of the machines to advance 
our understanding of complicated natural phenomena.

\section*{Acknowledgments}
We thank Christie Chiu, Annabelle Bohrdt, and Neil Switz for useful discussions.
E.K. acknowledges support from the National Science Foundation (NSF) under Grant 
No. DMR-1918572. Computations were performed in part on the Spartan high-performance 
computing facility at San Jos\'{e} State University, which is supported by the 
NSF under Grant No. OAC-1626645. The work of R.T.S. was supported by Grant No.
DE-SC0014671 funded by the U.S. Department of Energy, Office of Science. 
The work of E.G.-S., B.M.S., and W.S.B.  was  supported  by the NSF under Grant  No. 
DMR-1607277, the  David  and  Lucile Packard  Foundation  under Grant  No. 2016-65128,  
and the AFOSR Young Investigator Research Program under Grant No. FA9550-16-1-0269. 
J.C. acknowledges support from the Natural Sciences and Engineering Research Council 
of Canada (NSERC), the Shared Hierarchical Academic Research Computing Network (SHARCNET), 
Compute Canada, Google Quantum Research Award, and the Canada CIFAR AI chair program.

\appendix
%%%%%%%%%%%%%%%%%%%%%%%%%%%%%%%%%%%%%%%%%%%%%%%%%%%%%%%%%%%%%%%%%%%%
\section{The Fermi-Hubbard Model}
%%%%%%%%%%%%%%%%%%%%%%%%%%%%%%%%%%%%%%%%%%%%%%%%%%%%%%%%%%%%%%%%%%%%

The Hamiltonian for the 2D Fermi-Hubbard model in particle-hole symmetric form is expressed as
%\begin{widetext} 
\begin{eqnarray}
\hat{H}&=&-t\sum_{\left <{\bf i},{\bf j}\right >\, \sigma}
\left(\hat{c}^{\dagger}_{\bf i \, \sigma} 
\hat{c}^{\phantom{\dagger}}_{\bf j \, \sigma} 
+ \text{H.c.}\right)\\
&+& U\sum_{\bf i} \left(\hat{n}_{{\bf i}\uparrow}-\frac{1}{2}\right) \left(\hat{n}_{{\bf i}\downarrow}-\frac{1}{2}\right)\nonumber \\
&-& \mu \sum_{\bf i}(\hat{n}_{\bf i\uparrow}+\hat{n}_{\bf i\downarrow})\nonumber,
\label{eq:ham}
\end{eqnarray}
%\end{widetext} 
where $\hat{c}^{\dagger}_{{\bf i}\sigma}$ 
($\hat{c}^{\phantom{\dagger}}_{{\bf i}\sigma}$) 
creates (annihilates) a fermion with spin $\sigma$ on site ${\bf i}$, and 
$\hat{n}_{{\bf i}\sigma}=\hat{c}^{\dagger}_{{\bf i}\sigma} 
\hat{c}^{\phantom{\dagger}}_{{\bf i}\sigma}$ is the number operator. 
$\left <{\bf i},{\bf j}\right >$ denotes nearest neighbors on a square lattice, $U=8t$ is the strength of the onsite 
repulsive interaction in the numerical simulations, and $\mu$ is the chemical potential. $\mu=0$ corresponds 
to half filling, although density fluctuations around half filling exist in our grand canonical ensemble.
$t=1$ (also $\hbar=1$ and $k_B=1$) sets the energy scale. The spin correlation function is calculated 
as $C_s({\bf r})=\left<\hat{S}_{z,{\bf i}}\hat{S}_{z,{\bf i+r}} \right>$, 
where $\hat{S}_{z,{\bf i}}=\frac{1}{2}
(\hat{n}_{{\bf i}\uparrow}-\hat{n}_{{\bf i}\downarrow})$ and $\left <\dots \right >$ denotes the expectation value. 
The local moment correlation function is calculated as
$C_m({\bf r})=\left<\hat{m}^2_{z,{\bf i}}\ \hat{m}^2_{z,{\bf i+r}} \right>$, where 
$\hat{m}^2_{z,{\bf i}}=(\hat{n}_{{\bf i}\uparrow}-\hat{n}_{{\bf i}\downarrow})^2$.\\
%\vspace{-0.1in}

%%%%%%%%%%%%%%%%%%%%%%%%%%%%%%%%%%%%%%%%%%%%%%%%%%%%%%%%%%%%%%%%%%%%
\section{Training the Convolutional Neural Network}
%%%%%%%%%%%%%%%%%%%%%%%%%%%%%%%%%%%%%%%%%%%%%%%%%%%%%%%%%%%%%%%%%%%%

We implement our CNNs using TENSORFLOW~\cite{tensorflow}. 
The minimalistic design in Fig.~\ref{fig:sample}(c) we have adopted reflects 
the need to reduce the number of free parameters to avoid overfitting given 
the sizes of our data sets. To train, we assign a label, $Y$,
to each snapshot based on the temperature at which it is taken, or whether
it is real or fake. Each label is stored in the one-hot format, i.e., 
a binary array of two numbers, one of which is 1 and the other 0. The 
index for 1 indicates the category (high or low temperature, or real or fake) to which each 
snapshot belongs. Given an input image ${\bf X}$, the value arriving at the first 
of the two output neurons of the CNN shown in Fig.~\ref{fig:sample}(c), e.g., at 
the neuron we have associated in our labels to the low-temperature (or real) 
snapshots, is
\be
O_1^{out} ({\bf X}) = \sum_h O_h^{hid}({\bf X}) \times W_h^{out\ (1)} + b^{out\ (1)},
\ee
where the sum is over hidden neurons, 
\begin{widetext}
\be
O_h^{hid}({\bf X}) &=&\textrm{ReLU}\left(\left [\frac{1}{N_s}\sum_{stride: s} 
\textrm{ReLU}\left ({\bf W}^{filter}\cdot {\bf X}(s)+b^{filter}\right)\right ]\times W_h^{hid}+ b^{hid}_h\right ),
\ee
\end{widetext}
$N_s$ is the number of strides the filter takes around the image convolving 
with different sections, ${\bf W}^{filter}$ is the matrix of pixel values for the 
filter, ${\bf X}(s)$ is the matrix of pixel values for the section of the image the filter is 
convolving with in stride $s$, ReLU is the rectified linear unit activation function, and 
$b^{filter}$, $W_h^{hid}$, $b^{hid}_h$, $W_h^{out (1)}$, and $b^{out (1)}$  are numbers 
representing other weights and biases in the network. $O_1^{out}$, along with the value arriving 
at the second output layer $O_2^{out} ({\bf X})$, are then passed through the softmax activation 
function to obtain two probabilities as network outputs:
%\begin{widetext}
\be
&[&\textrm{Network Output}^{(1)}({\bf X}),\ \textrm{Network Output}^{(2)}({\bf X})]\nonumber\\
&=& \textrm{softmax} [O_1^{out}({\bf X}),\ O^{out}_2({\bf X})].
\label{eq:netout}
\ee
%\end{widetext}
The input snapshot is classified as belonging to category $i$ if $O^{net}_i$ is
the higher probability, where $O^{net}_i=\textrm{Network Output}^{(i)}$ for brevity. 
The accuracy is defined as the percentage of correct classifications 
given known labels $Y$. The convolution of the
trained filter with sections of the input image as it moves around in strides of one in 
every direction creates a ``feature map" in which large overlaps between patterns in the 
filter and the image are highlighted.

For training, we use the Adam optimizer, which is an extension of stochastic gradient 
descent, to minimize the cross-entropy cost function, defined as 
\be
\label{eq:loss}
c = &-&\frac{1}{N_d}\sum_{\bf X} \sum_{i=1}^2 ( Y_i({\bf X}) \ln [O^{net}_i({\bf X})]\\ 
&+& [1-Y_i({\bf X})] \ln [1-O^{net}_i({\bf X})] ),\nonumber
\ee
where $N_d$ is the number of data. 
During the training, we keep between 10\% and 20\% of the 
snapshots from the CNN and use them to perform unbiased testing 
of the accuracy.

%%%%%%%%%%%%%%%%%%%%%%%%%%%%%%%%%%%%%%%%%%%%%%%%%%%%%%%%%%%%%%%%%%%%
\subsection{CNN with more than one filter}
%%%%%%%%%%%%%%%%%%%%%%%%%%%%%%%%%%%%%%%%%%%%%%%%%%%%%%%%%%%%%%%%%%%%

In cases where we have more than one filter in the convolutional layer, 
we have modified the architecture to have no fully connected hidden layer in order 
to reduce the total number of network parameters; the output of each filter after 
pooling is instead fully connected to the output layer. The value arriving at the output neuron that is 
responsible for firing when a real snapshot ${\bf X}$ is provided to the input, $O_1^{out}({\bf X})$, 
can then be expressed as a linear combination of contributions from individual filters:
\begin{widetext}
\begin{eqnarray}
O_1^{out}({\bf X}) &=& \sum_{m=1}^{N_f} F_m^{(1)}({\bf X}),\\
F_m^{(1)}({\bf X})&=&\left [\frac{1}{N_s}\sum_{stride: s} \textrm{ReLU}\left
 ({\bf W}_m^{filter}\cdot {\bf X}(s)+b_m^{filter}\right)\times W_m^{out (1)}\right ]+\frac{b^{out (1)}}{N_{f}},
\label{eq:Fm}
\end{eqnarray}
\end{widetext}
where $N_f$ is the number of filters, and $W_m^{out (1)}$ and $b_m^{filter}$
are again numbers representing other weights and biases in the network. As in the case of 
the CNN with one filter, the network output is obtained using Eq.~(\ref{eq:netout}).

%%%%%%%%%%%%%%%%%%%%%%%%%%%%%%%%%%%%%%%%%%%%%%%%%%%%%%%%%%%%%%%%%%%%
\subsection{Effect of individual filters}
%%%%%%%%%%%%%%%%%%%%%%%%%%%%%%%%%%%%%%%%%%%%%%%%%%%%%%%%%%%%%%%%%%%%

To estimate the effect of filter $m$ on the outcome, we replace $O^{(1,2)}({\bf X})$ with 
$F_m^{(1,2)}({\bf X})$ before the softmax function,
\be
&[&\textrm{Network Output}_m^{(1)}({\bf X}),\textrm{Network Output}_m^{(2)}({\bf X})]\nonumber\\ 
&=&\textrm{softmax} [F_m^{(1)}({\bf X}), F_m^{(2)}({\bf X})],
\label{eq:netoutm}
\ee
so that we can interpret 
$\left [\textrm{Network Output}_m^{(1)} ({\bf X}^{real}) -\textrm{Network Output}_m^{(1)}  ({\bf X}^{fake})\right ]$
as the percentage the network output for ${\bf X}$, based on the action of filter $m$ alone, 
has to do with factors other than the density itself.

%%%%%%%%%%%%%%%%%%%%%%%%%%%%%%%%%%%%%%%%%%%%%%%%%%%%%%%%%%%%%%%%%%%%
\subsection{Augmentation of data}
%%%%%%%%%%%%%%%%%%%%%%%%%%%%%%%%%%%%%%%%%%%%%%%%%%%%%%%%%%%%%%%%%%%%

We augment~\cite{c_shorten_19} our data for the experimental single-species snapshots
before training [Figs.~\ref{fig:5x5}(a) and ~\ref{fig:5x5}(b)] 
by applying point-group symmetries of the square lattice to each 
snapshot. 
This will increase the data by a factor of 8, and in these cases, helps make
the training smoother and/or faster. However, we find that generally, such augmentation of data, or breaking the 
$20\times 20$ images into smaller subregions,
does not significantly affect the final accuracy. This is consistent
with the fact that the network 
architectures we have used in this work are not deep and do not show 
serious overfitting even before the data augmentation.

%%%%%%%%%%%%%%%%%%%%%%%%%%%%%%%%%%%%%%%%%%%%%%%%%%%%%%%%%%%%%%%%%%%%
\subsection{Training Progression}
%%%%%%%%%%%%%%%%%%%%%%%%%%%%%%%%%%%%%%%%%%%%%%%%%%%%%%%%%%%%%%%%%%%%

\begin{figure}
\centerline{\includegraphics*[width=3.3in]{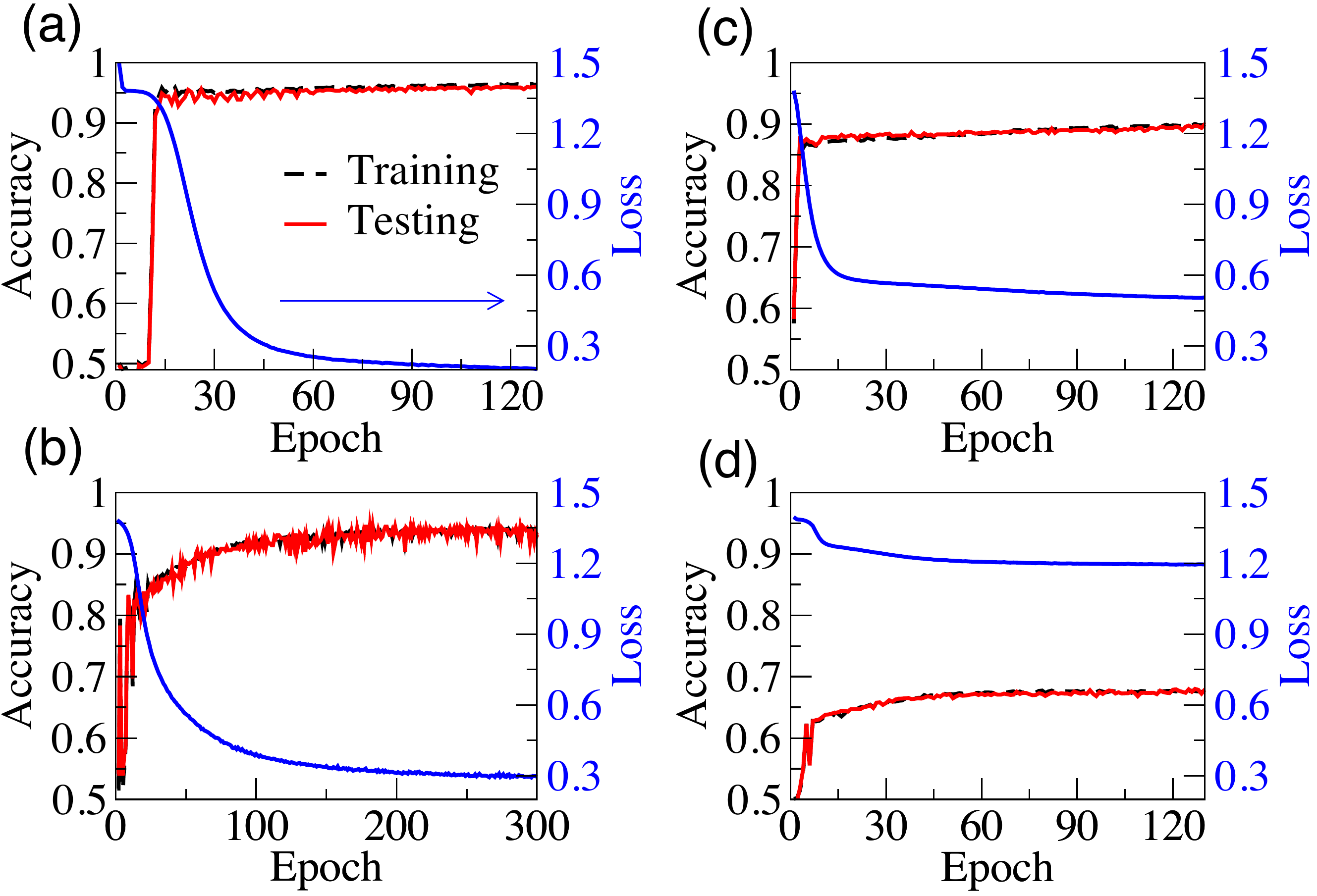}}
\caption{{\bf Progression of CNN accuracy during training with single-species 
snapshots.} (a) Training and unbiased testing accuracies, along with the 
value of the loss function when using experimental snapshots at $n\sim 1$ to 
train the CNN with one filter, leading to the filter shown in Fig.~\ref{fig:5x5}(a). 
(b-d) Same as (a), but for trainings leading to filters in 
Figs.~\ref{fig:5x5}(b)-\ref{fig:5x5}(d). DQMC snapshots are used in (c) and (d). (a) and (c)
correspond to trainings near half filling and (b) and (d) correspond to trainings 
at $n=0.82$.
\label{fig:progup}}
\end{figure}

\begin{figure}
\centerline{\includegraphics*[width=3.3in]{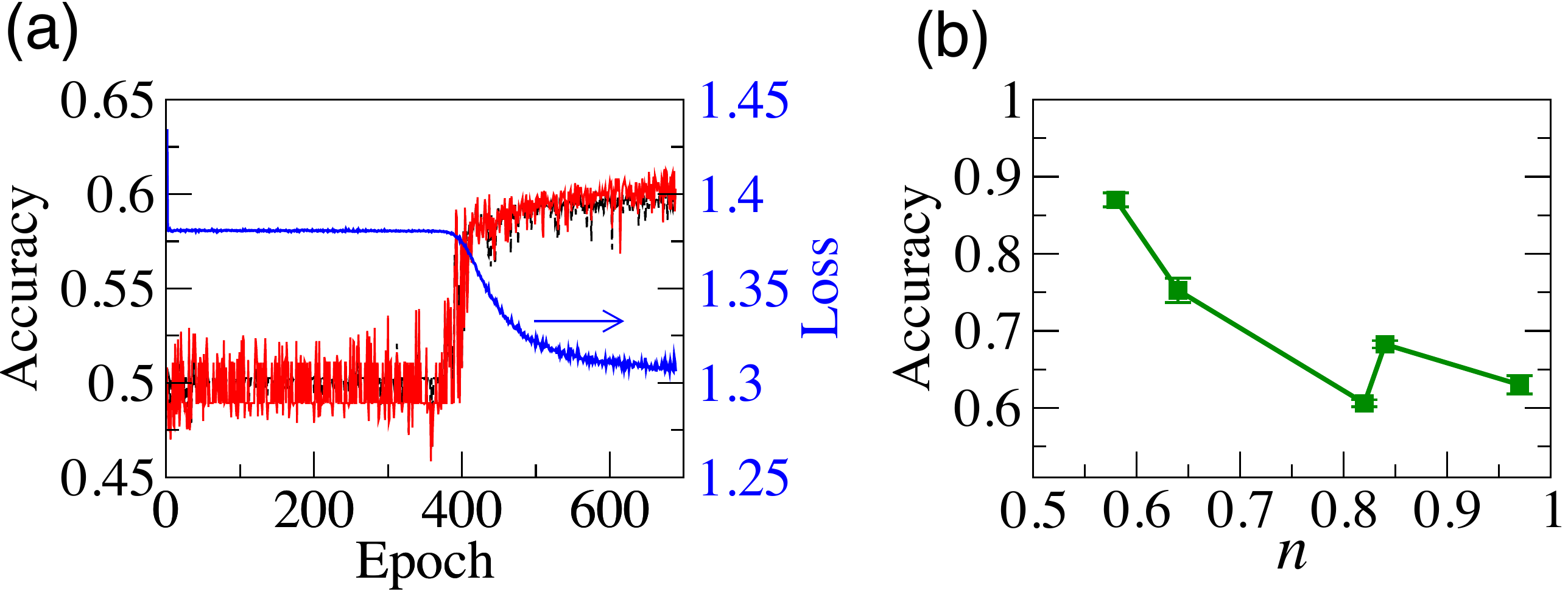}}
\caption{{\bf Progression of CNN accuracy during training with local 
moment snapshots.} (a) Same as Fig.~\ref{fig:progup}(a), but for the training
with snapshots of local moments at $n=0.82$ leading to the filter shown in 
Fig.~\ref{fig:ExptLM}(c). (b) The average of maximum testing accuracies, 
obtained through four or five separate trainings, vs density.
\label{fig:accn}}
\end{figure}

To monitor the training progression and look for signs of overfitting, 
especially in the case of CNNs with more than one filter, we track the 
training and the unbiased testing accuracies as well as the loss function,
defined in Eq.~(\ref{eq:loss}), over {\it epochs}. An epoch is 
when the network has gone over the entire data set once.

Figure~\ref{fig:progup} shows the results for various trainings using 
single-species snapshots that lead to 
filters presented in Fig.~\ref{fig:5x5}. In each case, we stop 
the training roughly when the loss function reaches a plateau. 
As can be seen, the training and the unbiased testing accuracies, 
which agree throughout the training process, also settle to their 
maxima.

Figure~\ref{fig:accn}(a) shows similar results for the sample training 
with snapshots of local moments that lead to the filter presented in 
Fig.~\ref{fig:ExptLM}(c). For this case, we find that it 
takes longer for any training to be achieved and that the fluctuations 
in the accuracies are larger. Figure~\ref{fig:accn}(b) shows the 
average of maximum accuracies achieved through four or five different training
runs with different random number seeds over the range of densities 
for which snapshots of local moments were available. We see that 
as the strength of local correlations decreases by increasing the 
density towards half filling [shown in Fig.~\ref{fig:ExptLM}(b)],
the maximum accuracies we can achieve also generally decrease. 

\begin{figure}
\centerline{\includegraphics*[width=3.3in]{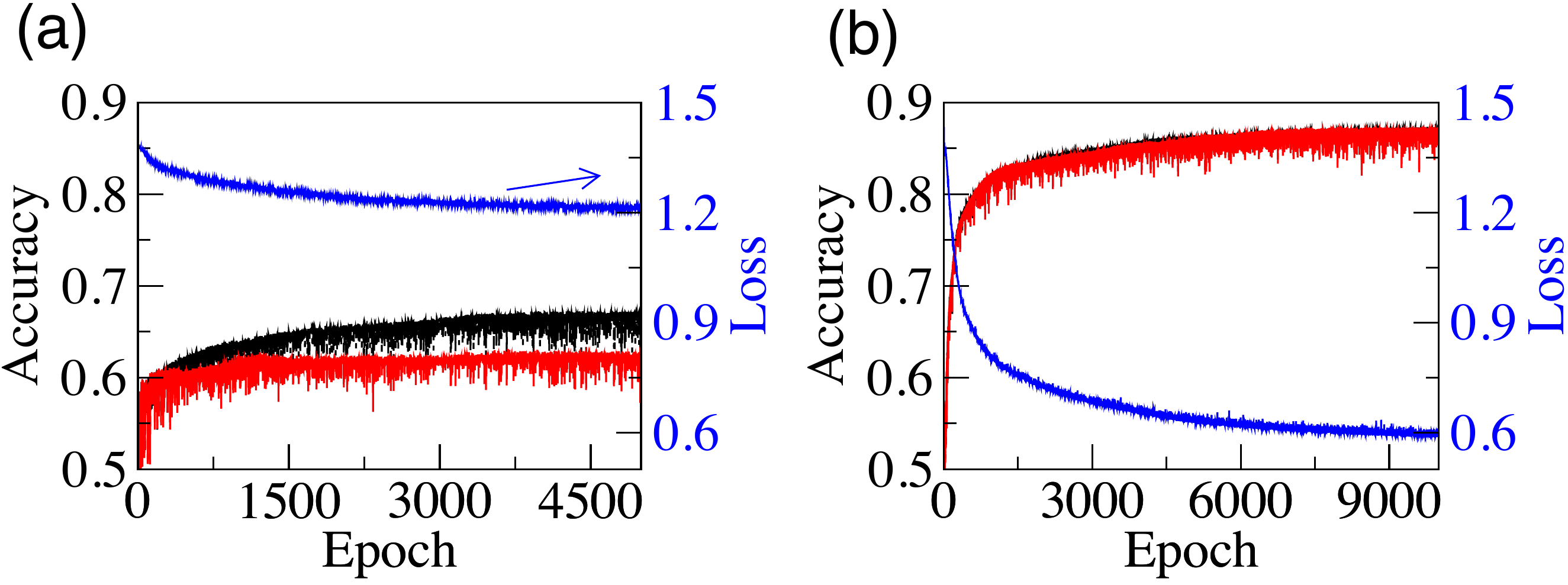}}
\caption{{\bf Progression of accuracy of CNNs with multiple filters during 
trainings with local moment snapshots.} Same as Fig.~\ref{fig:progup}, except 
that (a) experimental and (b) DQMC snapshots of local moments are 
used to train CNNs with 6 and 16 filters, showcased in Fig.~\ref{fig:Expt6filter}.
\label{fig:prog}}
\end{figure}

Figure~\ref{fig:prog} shows the same training progressions for the 
case of CNNs with multiple filters, trained on snapshots of local 
moments at $n=0.82$, leading to results shown in Fig.~\ref{fig:Expt6filter}. 
Despite the deviation of
the average of two accuracies from each other beyond $\sim 1000$ 
epochs when experimental snapshots are used [Fig.~\ref{fig:prog}(a)], 
signaling the beginning of overfitting due to the relatively
large number of free parameters in the CNN, large fluctuations 
in the accuracies cause overlapping of the two curves 
even after 5000 epochs. When DQMC snapshots are used, the signs
of overfitting are observed only after increasing the number of
$5\times 5$ filters to about 60.

%%%%%%%%%%%%%%%%%%%%%%%%%%%%%%%%%%%%%%%%%%%%%%%%%%%%%%%%%%%%%%%%%%%%
\section{CNN with Sixteen Filters for Theory Snapshots of Local Moments}
%%%%%%%%%%%%%%%%%%%%%%%%%%%%%%%%%%%%%%%%%%%%%%%%%%%%%%%%%%%%%%%%%%%%

\begin{figure*}[h!]
\centerline{\includegraphics*[width=0.9\textwidth]{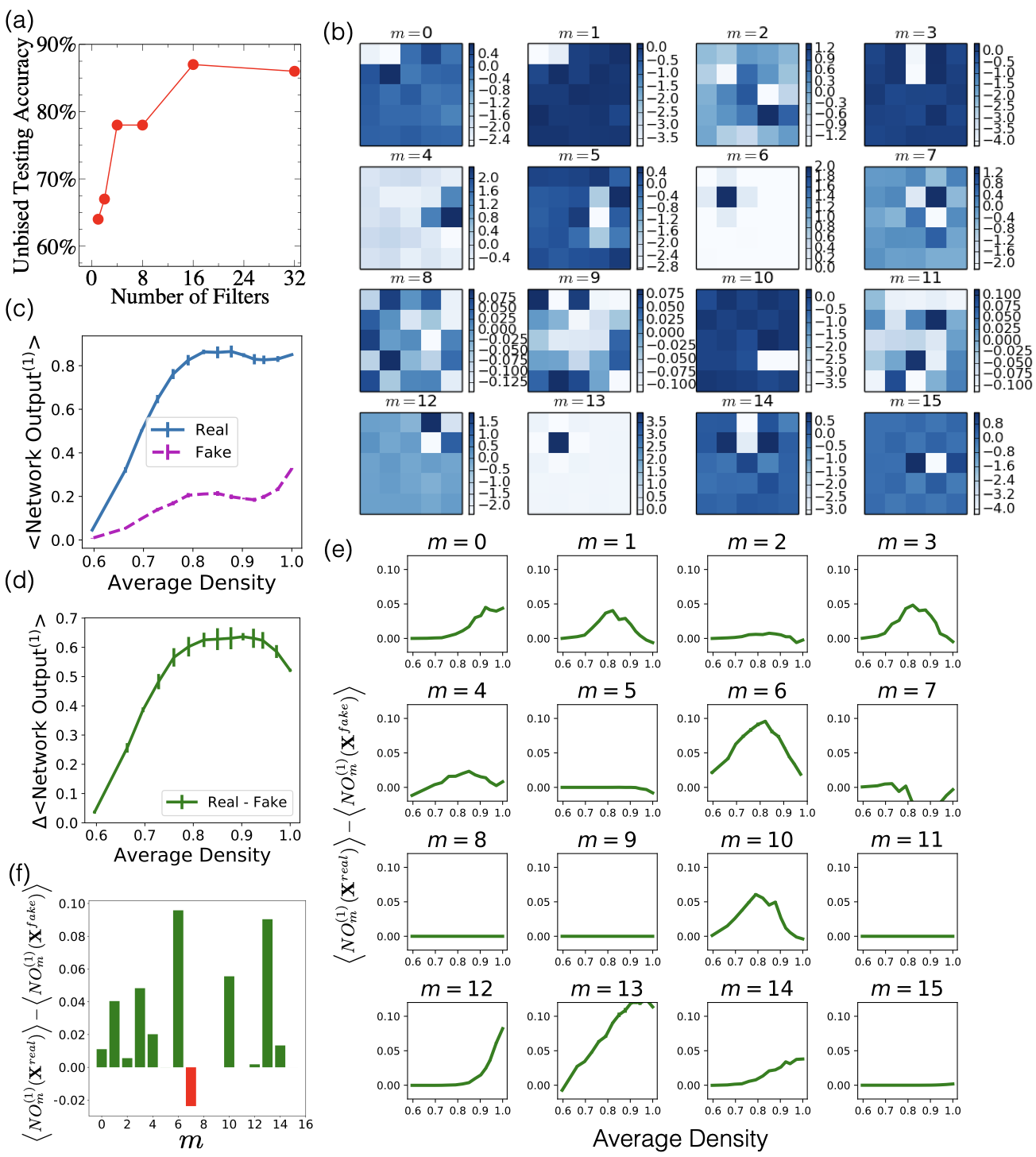}}
\caption{{\bf Analysis of DQMC local moment snapshots using a CNN with 16 filters.} 
(a) The general improvement in the prediction accuracy of the CNN by increasing the number of filters.
(b) Trained filters when $n=0.82$ and $T=0.44t$. 
Note that pixel values or their range 
in one filter should not be compared with those in other filters since a filter-dependent 
bias is added to the result of the convolution before it is passed through the ReLU 
activation function (see Appendix B). 
The testing accuracy is around $87\%$.
(c) Average network output when a real or fake DQMC 
snapshot is provided as input, as a function of average density for this CNN.
Here, 1 means all the snapshots are classified as likely to be real $n=0.82$ snapshots,
0 means all the snapshots are classified as likely to be fake $n=0.82$ snapshots, and 
0.5 means neither of the two options is preferred. 
(d) The difference between the two curves in (c), representing the average percentage the 
decisions made by the CNN have to do with factors other than the density itself. 
(e) Similar to (d) when the CNN has access to one filter at a time (see Appendix B). 
(f) Same as in (e), but at $n=0.82$.
\label{fig:NetOutLM}}
\end{figure*}

Figure~\ref{fig:NetOutLM} shows the result of training a CNN with 16 filters using 15000 $10\times 10$
theory snapshots of local moments. This is the same CNN whose filters are featured 
in Fig.~\ref{fig:Expt6filter}(d). To obtain the snapshots in the DQMC approach, we
note that for a particular auxiliary field, the expectation value of the local double 
occupancy reduces to its uncorrelated value; the product of the expectation values for spin-up and 
spin-down occupancies. Therefore, the local moment at site $i$ can be written as 
$\left <\hat{n}_{{\bf i}\uparrow}\right> + \left <\hat{n}_{{\bf i}\downarrow}\right> -2 
\left <\hat{n}_{{\bf i}\uparrow}\right>\left <\hat{n}_{{\bf i}\downarrow}\right>$.

Figure~\ref{fig:NetOutLM}(a) shows the general improvement of the testing accuracy by 
increasing the number of filters. Figure~\ref{fig:NetOutLM}(b) shows the 16 trained 
filters, which similarly to what we find for 
spin, point to only short-range fluctuations. One can find many redundancies. 
However, a few representative patterns [those features in Fig.~\ref{fig:Expt6filter}(d)] emerge. 
Figure~\ref{fig:NetOutLM}(c) shows the average network output
when the network trained at $n=0.82$ is tested on configurations across a 
range of densities. We also test the network on fake snapshots. A nonmonotonic 
behavior emerges in both cases. Subtracting the average network output for the real and fake
snapshots at each density results in a curve that has a broad peak around $n=0.85$ 
and showcases the extent of learned correlations between local moments in this CNN that 
have to do with factors other than the average density itself [Fig.~\ref{fig:NetOutLM}(d)]. 
See Appendix B for more details.

\clearpage

To attribute certain features seen in trained filters in Fig.~\ref{fig:NetOutLM}(b)
to correlations unique to the NFL region, we have to rule out their dominance at other densities. 
Figure~\ref{fig:NetOutLM}(e) shows the performance of individual filters 
over the same range of densities we used to study the network output.  
Figure~\ref{fig:NetOutLM}(f) further highlights the values in Fig.~\ref{fig:NetOutLM}(e) 
at $n=0.82$. Based on these results, filters that significantly contribute to the CNN's 
decision-making process and are unique to the NFL phase and, therefore, are the best 
candidates for offering insight into local moment fluctuations are $m=1, 3, 6, 7$, and $10$.
The most frequently seen correlation seems to be the one between two neighboring 
empty sites. Filters $m=6$ and $13$ signal that the network also partly uses the information about 
the density gradient near local moments to make a decision.

%%%%%%%%%%%%%%%%%%%%%%%%%%%%%%%%%%%%%%%%%%%%%%%%%%%%%%%%%%%%%%%%%%%%
\section{Determinantal Quantum Monte Carlo Snapshots}
%%%%%%%%%%%%%%%%%%%%%%%%%%%%%%%%%%%%%%%%%%%%%%%%%%%%%%%%%%%%%%%%%%%%

The implementation of DQMC used in
the work proceeds via the exact rewriting
of the interacting electron-electron problem as independent electrons
moving in a space-imaginary time auxiliary field $h({\bf r},\tau)$.
This reformulation involves first expressing the partition function
${\cal Z}$ for the original Hubbard Hamiltonian as a path integral, and
then the use of a Hubbard-Stratonovich transformation to decouple the
electrons.  The original fermionic degrees of freedom are then traced
out analytically, leaving an equivalent expression for ${\cal Z}$ as an
integral over $h({\bf r},\tau)$.  Detailed descriptions can be found in 
Refs.~\cite{r_blankenbecler_81,s_white_89,assaad08}.

\begin{figure}[t]
\centerline{\includegraphics*[width=3.3in]{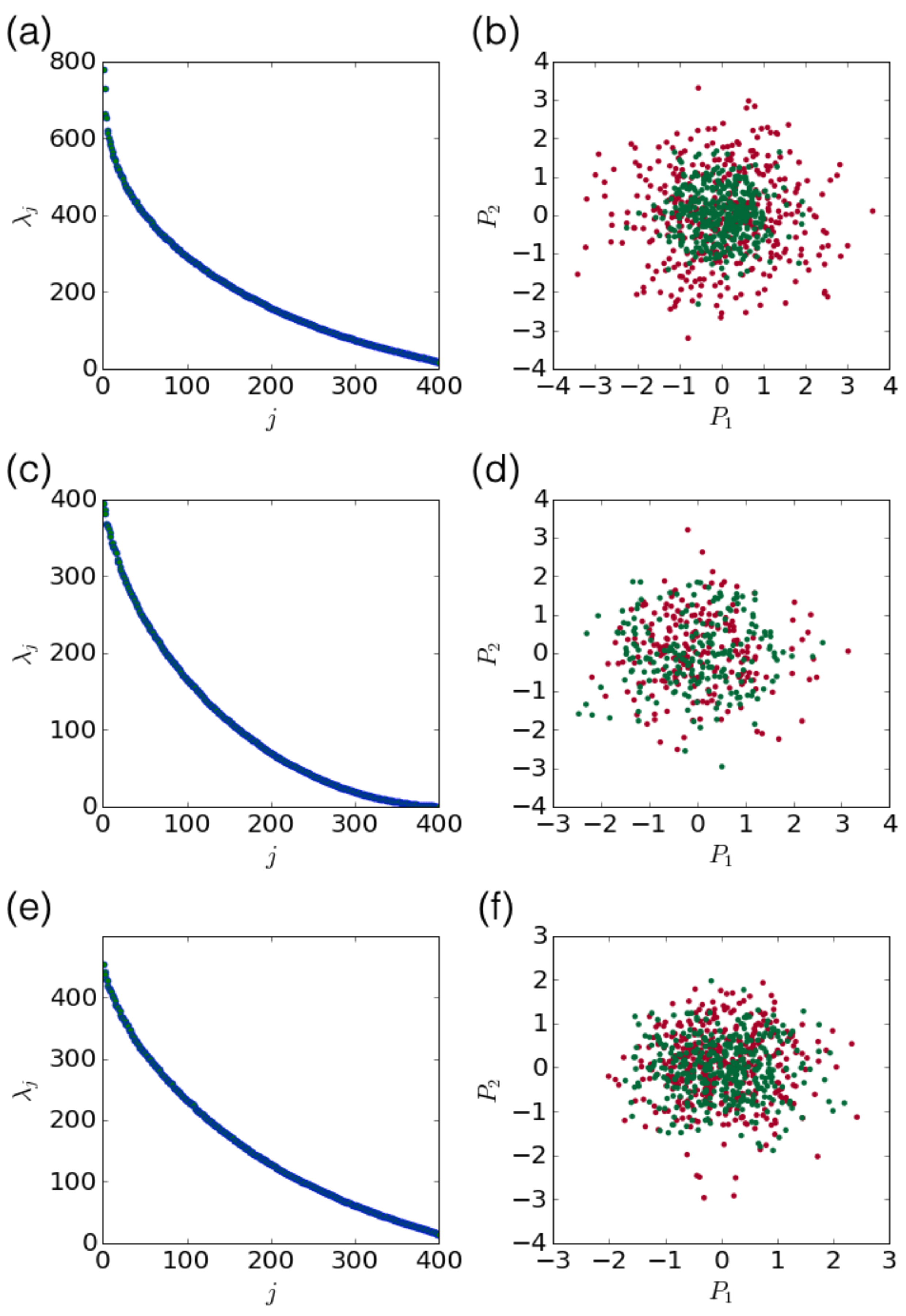}}
\caption{{\bf PCA of the experimental 
snapshots.}~(a, b) Eigenvalues of the covariance matrix of data 
for single-species snapshots at $n\sim 1$ and their projection to the 
space of the two largest principal components. The same 402 
snapshots at each of the two extreme temperatures, used for Fig.~\ref{fig:5x5}, are also used here. Although 
the data points corresponding to the lower temperature [green (dark gray) circles in 
(b)] are slightly less spread out than their high temperature 
counterparts [red (light gray) circles in (b)], there is no obvious clustering of
data based on temperature. 
(c, d) The same as (a) and (b), except for $n=0.82$ and that
216 available snapshots at each of the two extreme temperatures are used. 
(e, f) The same as (c) and (d), except that
400 real and 400 fake snapshots of local moments are used.
No discernible patterns emerge in these projections either.
\label{fig:pca}}
\end{figure}

Here we focus on aspects of DQMC which have specific implications to the
machine learning process.  The most crucial is that, unlike world-line
quantum Monte Carlo methods or cold-atom experiments, which directly sample the $0$ or $1$
occupation of sites ${\bf r}$ by the fermions, at any point (snapshot)
in a DQMC simulation, the fermionic occupation is represented by a real
number giving the probability of occupation of that site in the specific
$h({\bf r},\tau)$ currently being sampled.  
As the temperature is lowered below $t$, sharper images containing pixels more 
closely resembling binary pixels in the experimental snapshots emerge. 
In fact, one can show that in the atomic limit, local expectation values 
approach step functions as $T\to 0$. This ``smearing" of the
occupation makes some aspects of machine learning via these
snapshots more challenging.  However, whether individual snapshots
present (0,1) fermion occupations or not, the strong correlation physics
(magnetism, superconductivity, strange metallicity) of the Hubbard model
needs to be built up from many thousands of snapshots. It is the task of
uncovering these many-body effects that is shared by the theoretical
and experimental images investigated here with AI.

The presence of the fermion ``sign problem"~\cite{e_loh_90,m_troyer_05,v_iglovikov_15}
in DQMC simulations can complicate the interpretation of the theory snapshots. 
To avoid this complication to the extent possible, at $n=0.82$, we work at temperatures 
where the problem is not severe, where at least $90\%$ of the auxiliary field 
configurations have a positive sign.
We also treat the network output during the testing process the same way we treat the 
expectation value of a conventional observable, $O$, by computing
$\left<O\mathcal{S}\right>/\left<\mathcal{S} \right>$ in place 
of $\left<O \right>$, where $\mathcal{S}$ is the sign associated with 
the auxiliary field configuration resulting in a snapshot~\cite{k_chng_17}. Like 
the network output, $O$ is often a nonlinear function of the auxiliary field.

%\clearpage

%%%%%%%%%%%%%%%%%%%%%%%%%%%%%%%%%%%%%%%%%%%%%%%%%%%%%%%%%%%%%%%%%%%%
\section{Principal Component Analysis}
%%%%%%%%%%%%%%%%%%%%%%%%%%%%%%%%%%%%%%%%%%%%%%%%%%%%%%%%%%%%%%%%%%%%

We have employed the linear principal component 
analysis (PCA)~\cite{pca}, an unsupervised learning algorithm, to perform 
dimensional reduction on the experimental data and examine whether any features
emerge in the low-dimensional space, revealing any potential linear relation  
between pixels as an indicator for identifying low-temperature snapshots.

In the PCA, one forms a {\it matrix of data}, ${\bf x}$ by flattening the matrix of 
pixel values for each snapshot and placing them as an array of $400$ 
binary numbers in each row (of ${\bf x}$). In the next step, the covariance matrix of
data is composed as ${\bf x}^T\cdot {\bf x}$. Diagonalizing the $400\times 400$ matrix, 
we obtain its eigenvalues, whose magnitudes are a measure for the variance of the data 
along the {\it principal axes}, determined by the corresponding eigenvectors. 
As demonstrated for the two-dimensional Ising model in a pioneering application 
in physics~\cite{Wang2016}, a dominant eigenvalue is indicative of a clear 
distinguishing pattern in the snapshots that can be represented as a linear combination 
of its pixels (their projection to the corresponding principal axis).

Figure~\ref{fig:pca} shows our results for both single-species ($n\sim 1$ and $n=0.82$) 
and local moment ($n=0.82$) snapshots. The PCA does not seem to be able to draw any 
particular distinction between low and high temperature or real and fake 
snapshots in any of the cases as there are no clear signs of clustering 
of data points based on temperature or whether or not they correspond to real snapshots. 
Figures~\ref{fig:pca}(a) and \ref{fig:pca}(b) show, respectively, the eigenvalues of the covariance matrix
of data for snapshots of single species near half filling and their projections to the 
space formed by the first two principal axes, representing
the directions of the largest variance in data. A large gap between 
the first few eigenvalues and the rest of them would indicate that there are clear 
linear indicators for distinguishing sets of data~\cite{Wang2016}, something we 
do not observe here. This is also inferred from the projection of data in Fig.~\ref{fig:pca}(b),
where, other than a slightly larger spread of data points at the lower temperature,
no clear separation between the hot and cold data points is formed in the space
of the first two principal components. Figures~\ref{fig:pca}(c)-\ref{fig:pca}(f) show a similar 
trend for both the single species and local moment snapshots at $n=0.82$.

%\bibliography{scibib}


\end{document}